\documentclass[12pt]{iopart}
\begin{document}
\title{Neutrons transition densities for the $2^+-8^+$ multiplet of states
in $^{90}$Zr }
\author{ M S Onegin\dag~and A V Plavko\ddag }

\address{\dag\

Theory Division, PNPI, Gatchina, Leningrad district, Russia
188350}

\address{\ddag\ St Petersburg State Polytechnic University, St
Petersburg, Russia 195251}
\ead{onegin@thd.pnpi.spb.ru}

\begin{abstract}

The neutron transition densities of the $2^+-8^+$ levels in $^{90}$Zr
were extracted in the process of analysing ({\bf p},p') scattering at
400 Mev. Its comparison with the proton transition densities for
these levels was undertaken. The radial shapes of the experimental
neutron and proton transition densities for each state were found
to be different.

\end{abstract}

NUCLEAR REACTIONS $^{90}$Zr (polarized p,p'), E = 400 MeV;
calculated $\sigma$ ($\theta$) and A$_y(\theta)$. Experimental Neutron
transition densities for low-lying positive-parity
excitations.

\submitto{\JPG}
\pacno{24.10.Eq;24.10.Ht;25.40.Ep;21.10.-k}


\nosections

The Nuclear structure of $^{90}$Zr has been analysed in proton
inelastic scattering in various publications
(see~\cite{Gazally,Baker,Lee} and references there for earlier
papers). In these works, the authors tried to describe
experimental data by the shell-model procedure with a limited
basis size or by the collective model of inelastic excitation. The
lack of the necessary transition strength in the microscopic
calculations made them introduce enhancement factors needed to
adjust the calculated cross-sections to experiment.

\begin{figure}
\begin{center}
\epsfxsize=14.8cm
\epsfysize=10.3cm
\epsfbox{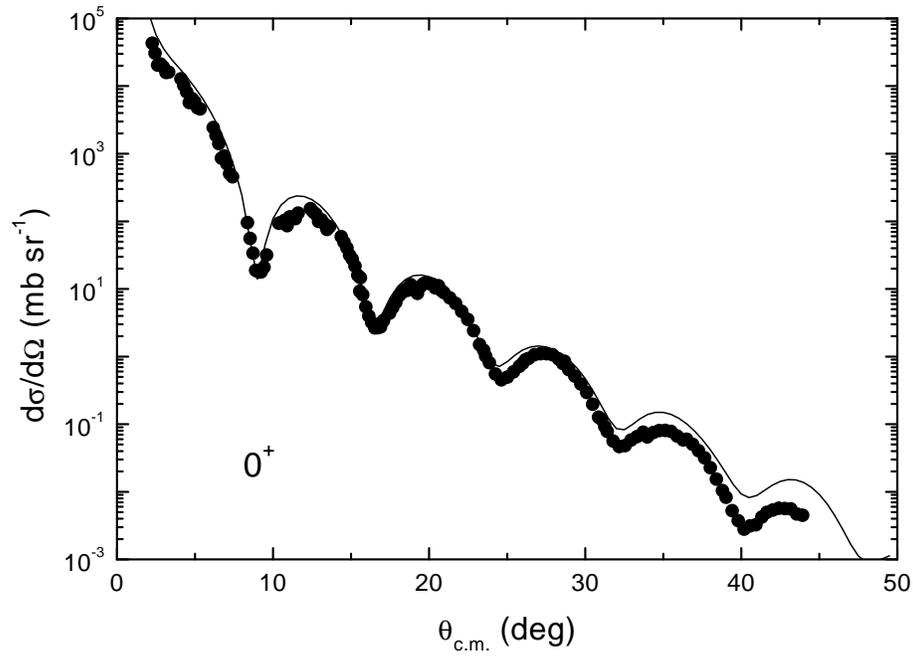}
\caption{\label{Fsel}Elastic cross section for proton scattering
from $^{90}$Zr at 400 MeV. The circles are the data
from Ref.~\cite{Lee}. The curve is a
microscopical folding model calculations.}

\end{center}
\end{figure}

\begin{figure}
\begin{center}
\epsfxsize=10.3cm
\epsfysize=14.8cm
\epsfbox{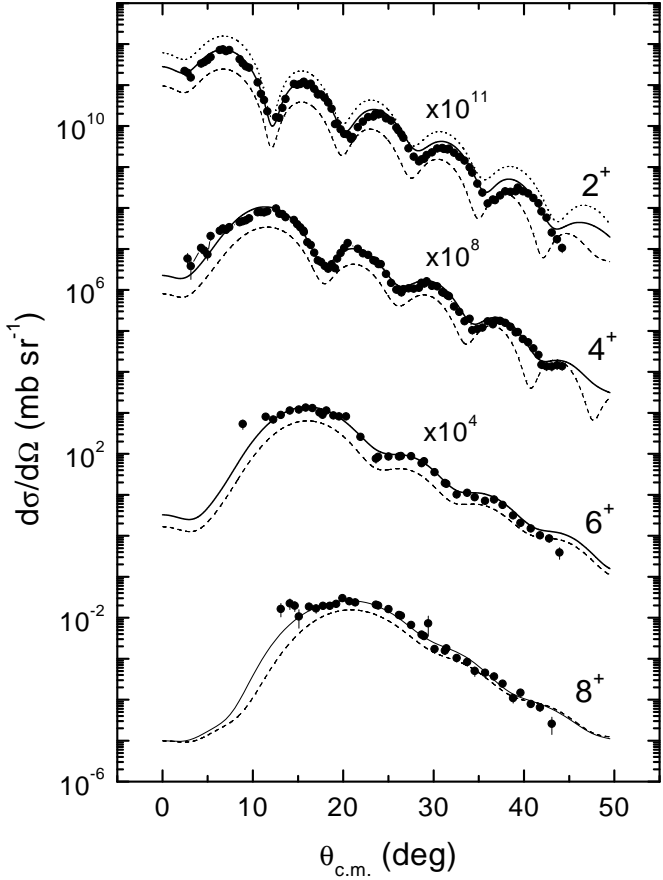}
\caption{\label{Sin}Inelastic cross sections for proton scattering
from
$^{90}$Zr at 400 MeV. The circles are the data from Refs.~\cite{Lee,Lee2}.
The differential cross sections are scaled by factors of $10^{11}, 10^8,
10^4$ for $2^+, 4^+, 6^+$ states, respectively, while $8^+$ results are
unscaled. The curves is a microscopical folding model calculations. The
dash curves - isoscalar option for transition densities; the full curves -
empirical fitted neutron transition densities; the dot curves - neutron
transition density from~\cite{Bartlett} }
\end{center}
\end{figure}

In the present paper, we employ a semi-microscopic approach
in which only  the matter component of transition densities is
used to describe the cross-section and analysing power of
inelastic scattering. Earlier ~\cite{Plavko,Kelly3,Plavko2} it was
demonstrated that this approach was adequate for the description
of inelastic scattering at medium and intermediate energies for
low-lying states with a large admixture of a core-polarisation
component in their nuclear wave functions. Since the proton
transition density can be determined independently in an electron
scattering experiment, proton scattering can be used to obtain the
second component of the matter density - the experimental neutron
transition density. In this letter, we use the experimental data
of ~\cite{Lee} to deduce these important characteristics of
nuclear excitation  for the whole multiplet of the
$2_1^+,4_1^+,6_1^+,8_1^+$   levels in $^{90}Zr$. Earlier a
slightly similar procedure was undertaken to deduce the model-
dependent neutron transition density. It was done for the first
member of this multiplet only ($2^+$). The result will be examined in
our letter further on.
In our study of proton scattering, we have  used the
calculation scheme provided by the linear expansion analysis (LEA)
code from Kelly~\cite{Kelly}. Our calculations have been performed
in the DWBA framework.  The same density-dependent nucleon-nucleon
interaction ~\cite{Geramb} is used in the calculations of the
optical potential and transition potential for inelastic
scattering in the folding-model formalism.  These potentials are
folded with the nucleon densities of the ground state ~\cite{Ray}
and the transition densities of the excited states, respectively.
We employ a zero-range approximation for exchange and use the
local density approximation based upon the density at the
projectile position in the analysis of a 400 MeV proton scattering
experiment. The comparison of the calculated elastic cross-
sections with the experimental data is presented in
Fig.~\ref{Fsel}. The overall agreement is satisfactory. Therefore,
we can be confident of the adequate use of the  nucleon-nucleon g-
matrix interaction and folding procedure at the given energy. For
the description of inelastic excitations only matter transition
densities are used in the folding procedure to obtain scattering
potentials. The proton transition densities have been obtained by
unfolding the proton charge densities from the charge transition
densities extracted in inelastic electron scattering
~\cite{Heisenberg}.
According to~\cite{Kelly}, neutron transition densities $\rho ^n_{tr}$ for
a transition of multipolarity L have been parametrised using the
Laguerre-Gaussian expansion (LGE)
$$\rho ^n_{tr,L}(r)=\sum_{\nu} a_{\nu} x^L e^{-x^2} L_{\nu}^k (2
x^2),$$
where $k=L+\frac{1}{2}$ and $x=r/b$. The fitting procedure is
similar to that employed in ~\cite{Kelly2} for the $^{88}Sr$
nucleus. The oscillator parameter $b$ has been set to 2.2 fm. $L_{\nu}^k$
is a generalised Laguerre polynomial of order $\nu$. The unknown
coefficients $a_{\nu}$ have been obtained by fitting the calculated
differential cross-sections for the nuclear levels in question to the
experimental data of ~\cite{Lee,Lee2}. The analysis includes a high-$q$
bias and an estimate of the incompleteness error that results from
limitation of the data to finite momentum transfer. A tail bias is used to
damp unphysical oscillations of the density for $r\ge r_m$, where it is
assumed that $\rho \propto e^{-dr}$ is beyond the match radius $r_m=6.5\,
fm$.  The parameter $d$ is adjusted to the fitted density at the match
radius $r_m$.  The fits to the cross-section data are displayed in
Fig.~\ref{Sin}.  The LGE expansion coefficients for the neutron transition
densities in question are tabulated in Table~\ref{ntd}. The fitted neutron
transition densities presented in Fig.3 are compared with proton transition
densities for the same levels.  In Fig.~\ref{Sin} the dash curve also
represents the calculated cross-sections in the approximation of a pure
isoscalar character of the excitations. It can be seen there from the
comparison with experiment that this approximation is unacceptable.
The transition strength is customarily characterised by the moment
$$ M_\lambda = \int dr r^{L+2} \rho ^\lambda _L (r), $$
where $\lambda$ =\,p\,or\,n for
protons and neutrons, respectively, and $\rho ^\lambda _L (r)$ is the
radial dependence of the corresponding transition density. The $M_\lambda$
value is highly sensitive to the tail bias of the radial density
distribution.  The latter, in its turn, is determined by small momentum
transfer experimental data for the transition analysed. Unfortunately, we
have a lack of such data for the $6^+$ and especially for the
$8^+$ states~\cite{Lee,Lee2}. Therefore, the extracted neutron transition
densities may be inaccurate for high multipolarities. The obtained
results $M_n /M_p $ for multipolarity L are presented in
Table~\ref{ntd}.
In Fig.4 we also present comparisons between the calculated
analysing powers and the corresponding experimental data for the
considered transitions. It can be seen that the use of the fitted
neutron transition densities improves the phase structure of the
analysing powers for the $2^+$ and $4^+$ states at large
scattering angles, though it slightly reduces the calculated
values in the region of 30 degrees in comparison with the
calculations using isoscalar transition densities for these
states. The overall agreement with experiment is nearly the same
for the both options of the transition densities. As a result, the
radial dependence of the neutron transition density has a small
influence upon the analysing powers; consequently, the analysing
power data have not been included in the fitting procedure in our
paper as well as in~\cite{Bartlett}.
The surface lobes in the  extracted neutron transition
densities for the $2^+-6^+$ states are greater in absolute value
than those in the proton densities; however, they are shifted to
the interior of the nucleus. That is why the ratio of the
transition strengths $M_n/M_p$ is smaller than a value of 1.0 (see
Table~\ref{ntd}). A comparable shifting of the surface lobes is
also observed for the $8^+$ level. However, the ratio between the
peaks of the two transition densities is reverse here, as compared
with the above mentioned states.  The application of our
transition densities, including that for the $8^+$, will be reported
in a separate paper.
Earlier the model-dependent neutron transition density for
the $2_1^+$ level was extracted ~\cite{Bartlett} in the
description of inelastic proton scattering at 500 Mev. Besides,
inelastic scattering of $^6$Li ions was used ~\cite{Sachler} to
extract the ratio $M_n /M_p$ and to test the transition densities
of this level, obtained from open-shell random phase approximation
(RPA) calculations. The value of $M_n /M_p$ obtained in
~\cite{Sachler} is $0.85\pm 0.10$ and in a special fit there is
$0.72\pm 0.10$, which agrees with our value $0.77\pm 0.03$.
However, the value obtained in ~\cite{Bartlett} ($M_n /M_p =1.47$)
considerably  deviates from ours. The radial dependence of the
neutron transition density from ~\cite{Bartlett}  for this state
is also presented in Fig.3 in comparison with that extracted in
the present paper. Their fitted result is slightly shifted to the
exterior of the nucleus, as compared with our experimental neutron
transition density, and it does not have any inner structure. When
we used the transition density of ~\cite{Bartlett} in our
calculations, we overestimated the differential cross-section for
this level (see Fig.~\ref{Sin}).
The deficiency exhibited by the transition densities of the
microscopic shell-model calculation (valence protons) is often
remedied by the coherent addition of a phenomenological core
vibration amplitude (see~\cite{Gazally,Lee} e.g.). However, our
results (Fig.  3) clearly demonstrate that such assumptions present too
naive a picture of $^{90}$Zr. Furthermore, our radial densities provide much
more insight into the structure of a transition than other predictions,
especially into the interior of the nucleus.  The proton transition
densities for the $2^+$ to $8^+$ multiplet arise from the same dominant
configuration  , and thus the shape of all these densities is determined
likewise by the shape of the   radial wave function. The neutron shell is
closed in $^{90}$Zr, but, as is seen from Fig. 3, the contributions of
neutron excitations are not weak and mostly outmeasure the proton
contributions.  Although the shapes of both contributions are different
across-the-board and, consequently, all the analysed excitations are far
from isoscalar, contrary to many common assumptions.  We have also used the
obtained transition densities for inelastic proton scattering at 61.2 MeV
and 800 MeV. The description appears to be rather good. Our demonstration
of the energy independence of the extracted neutron densities confirms the
accuracy of the analysis procedures.

\ack

\begin{table}
\caption{\label{ntd} Neutron transition densities expansion
coefficients
$a_\nu$ for $^{90}Zr$ expressed in units $fm^{-3}$.}
\lineup
\begin{indented}
\item[]\begin{tabular}{@{}lll}
\br
$\nu$ & $2_1^+$ & $4_1^+$  \\
\mr
1 &$(\m 7.96\pm 3.73)\times 10^{-3}$ &$(\m 5.52\pm 0.84)\times
10^{-3}$ \\
2 &$(-1.13\pm 0.15)\times 10^{-2}$ &$(-2.41\pm 0.18)\times 10^{-
3}$ \\
3 &$(\m 1.03\pm 0.06)\times 10^{-2}$ &$(\m 2.29\pm 0.77)\times
10^{-4}$ \\
4 &$(\m 2.88\pm 0.32)\times 10^{-3}$ &$(\m 1.86\pm 0.25)\times
10^{-4}$ \\
5 &$(-0.40\pm 1.53)\times 10^{-4}$ &$(\m 1.64\pm 0.14)\times 10^{-
4}$ \\
6 &$(-4.79\pm 0.33)\times 10^{-4}$ &$(\m 2.43\pm 0.79)\times 10^{-
5}$ \\
$M_n /M_p$ & $0.77\pm 0.03$ & $0.89\pm 0.05$ \\
\mr
$\nu$ & $6_1^+$ & $8_1^+$ \\
\mr
1 & $(\m 1.21\pm 0.07)\times 10^{-3}$ &$(\m 1.08\pm 0.08)\times
10^{-4}$ \\
2 & $(-4.42\pm 2.11)\times 10^{-5}$ &$(\m 1.21\pm 0.34)\times
10^{-5}$ \\
3 & $(-2.18\pm 0.59)\times 10^{-5}$ &$(\m 5.63\pm 1.10)\times
10^{-6}$ \\
4 & $(-1.47\pm 0.28)\times 10^{-5}$ &$(\m 5.84\pm 2.03)\times
10^{-7}$ \\
5 & $(-9.62\pm 8.46)\times 10^{-7}$ &$(\m 1.70\pm 0.72)\times
10^{-7}$ \\
6 & $(-1.17\pm 0.26)\times 10^{-6}$ &$(-1.89\pm 0.27)\times 10^{-
7}$ \\
$M_n /M_p$ & $0.71\pm 0.05$ & $1.43\pm 0.15$ \\
\br
\end{tabular}
\end{indented}
\end{table}

\input epsf
\newpage

\clearpage
\newpage
\begin{center}
\begin{picture}(500,400)
\put(0,200) {
\epsfxsize=9.9cm
\epsfysize=6.9cm
\epsfbox{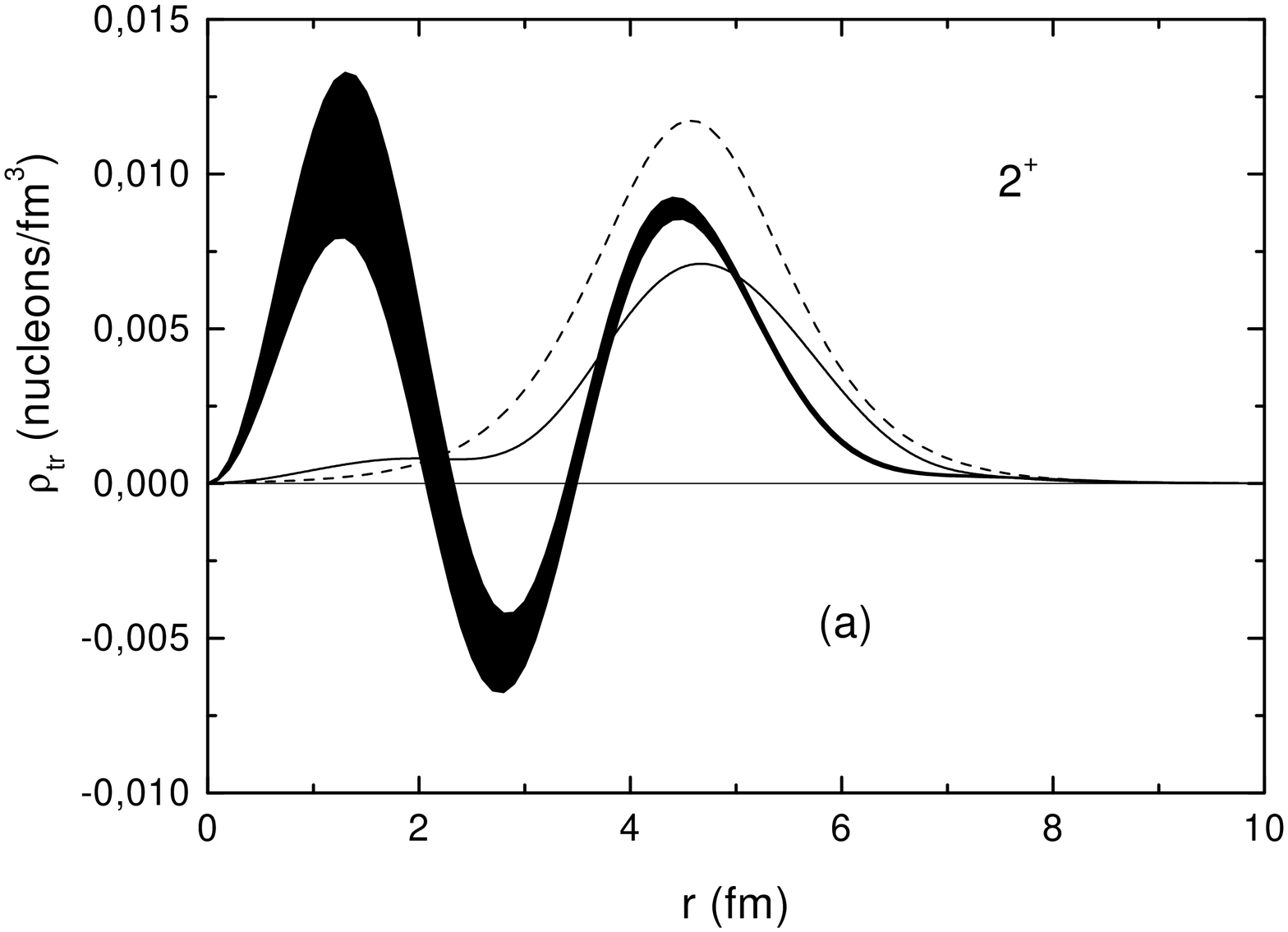}}
\put(250,200) {
\epsfxsize=9.9cm
\epsfysize=6.9cm
\epsfbox{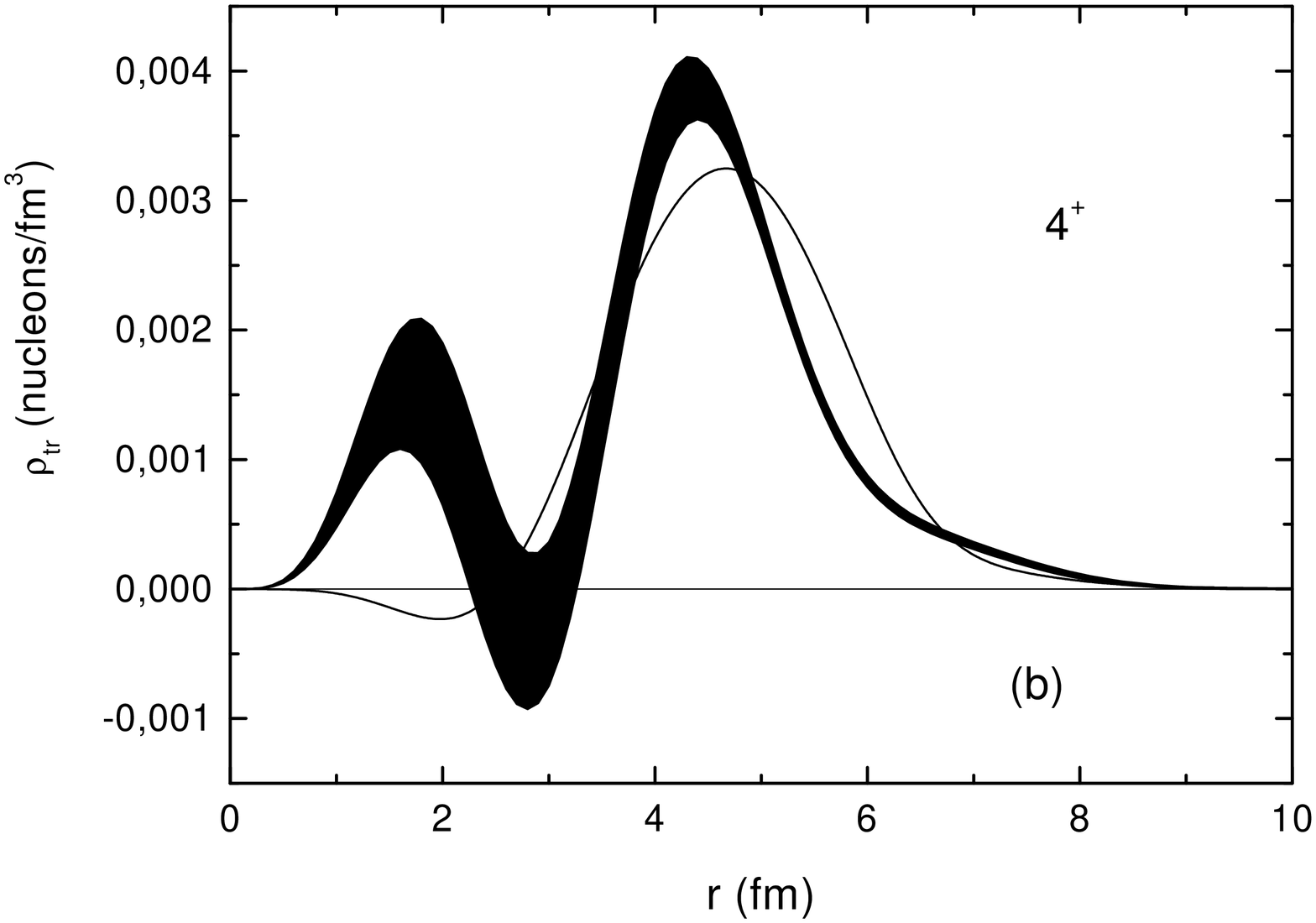}}
\put(0,0) {
\epsfxsize=9.9cm
\epsfysize=6.9cm
\epsfbox{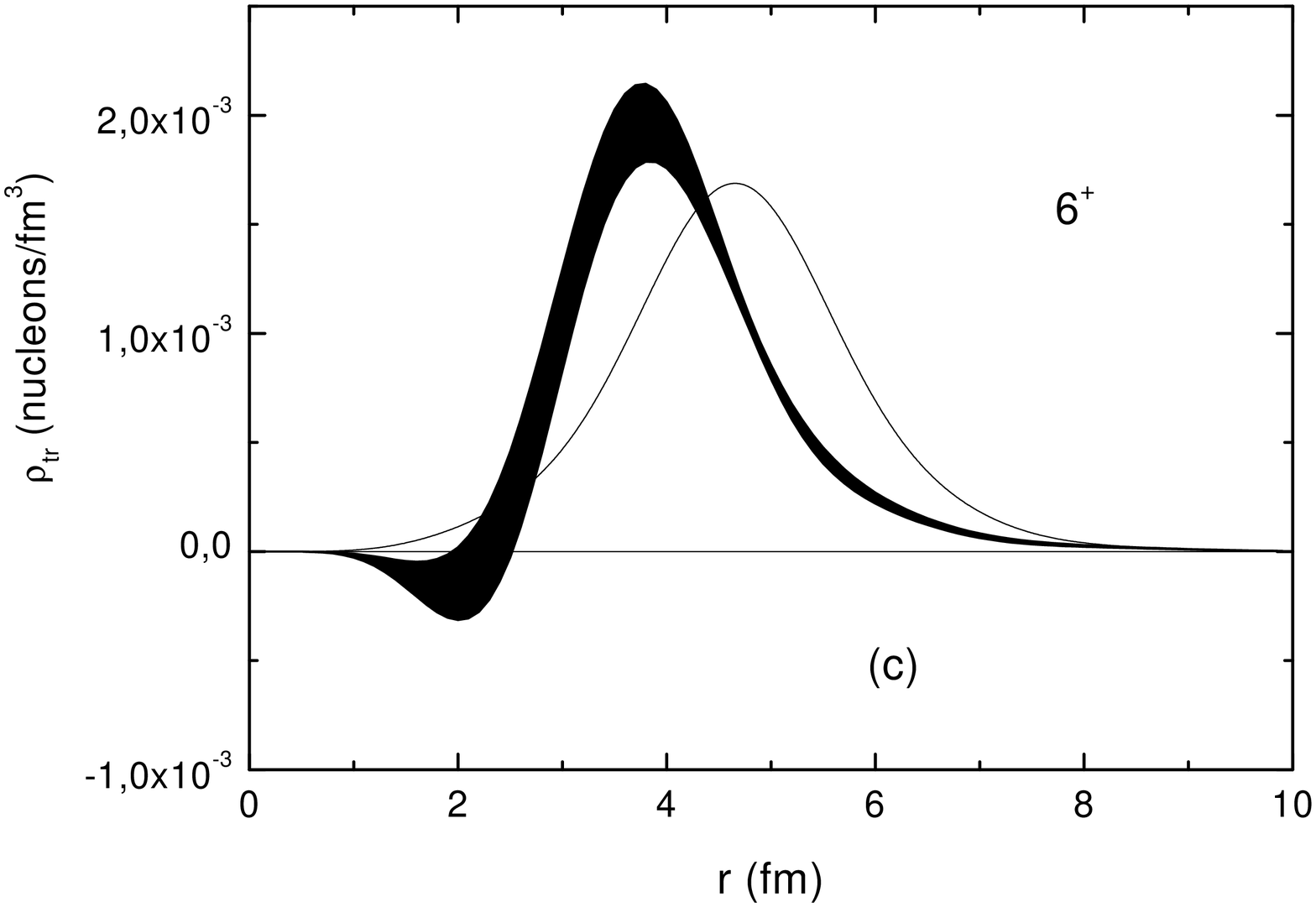}}
\put(250,0) {
\epsfxsize=9.9cm
\epsfysize=6.9cm
\epsfbox{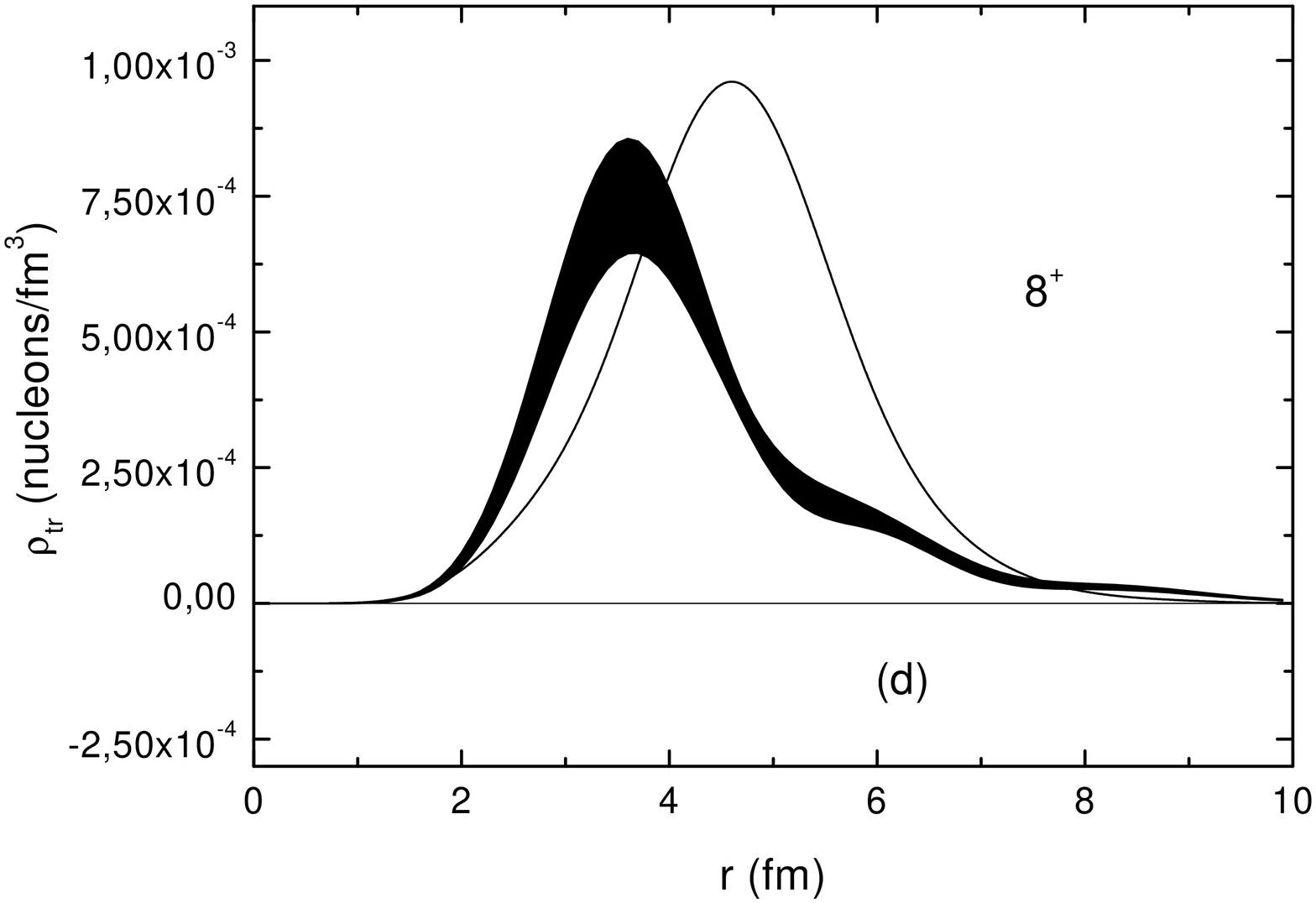}}
\end{picture}

Fig.3. The experimental neutron transition densities for $2_1^+$
(a); $4_1^+$ (b); $6_1^+$ (c); $8_1^+$ (d) (bands) are
compared with
point-proton transition densities (solid curve) unfolded from the
(e,e')
results~\cite{Heisenberg}. In part (a) of the figure
neutron transition density from~\cite{Bartlett} for $2_1^+$ state
is
present (dash curve).
\end{center}

\newpage
\begin{center}
\begin{picture}(500,400)
\put(0,200) {
\epsfxsize=9.9cm
\epsfysize=6.9cm
\epsfbox{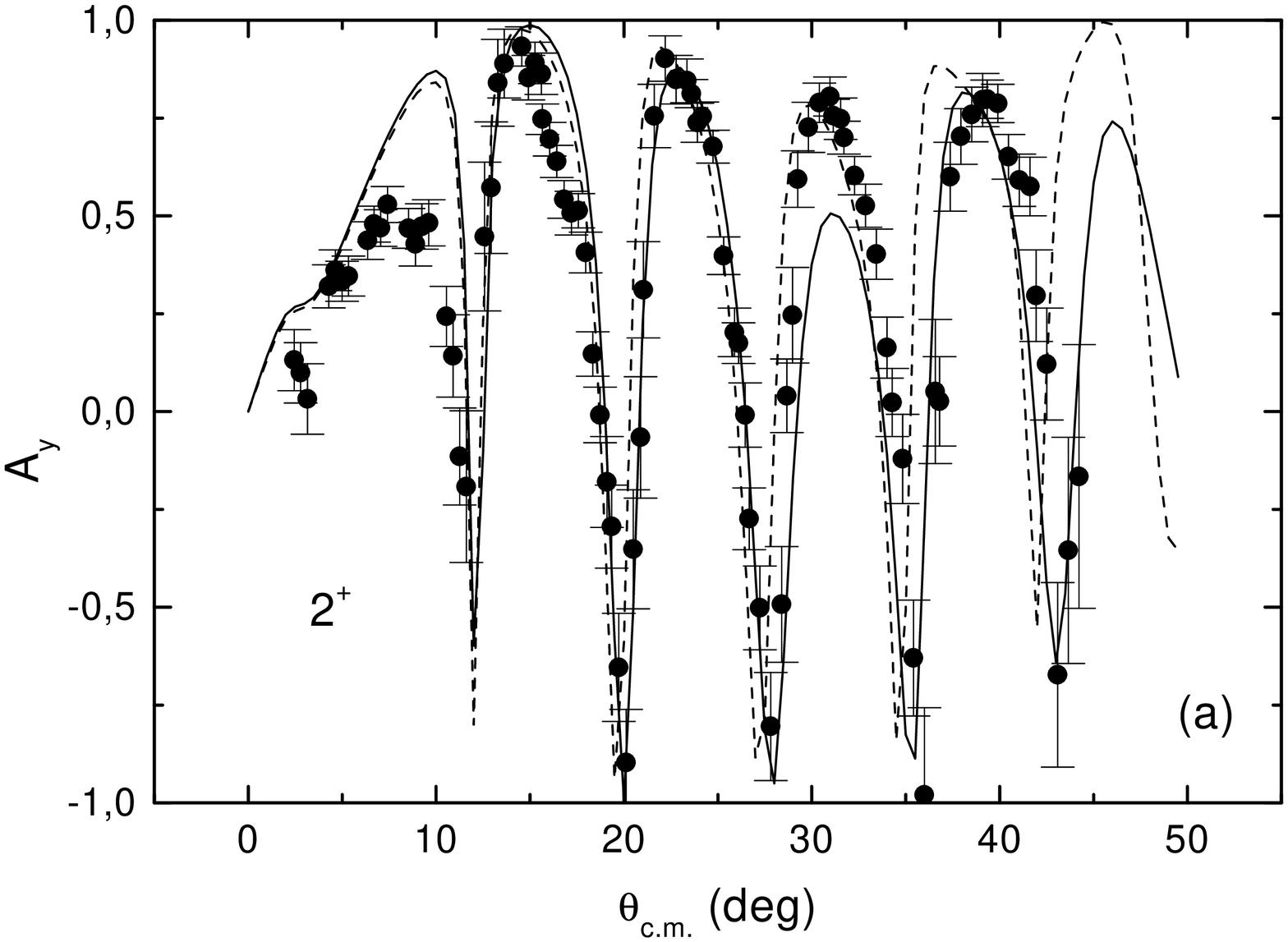}}
\put(250,200) {
\epsfxsize=9.9cm
\epsfysize=6.9cm
\epsfbox{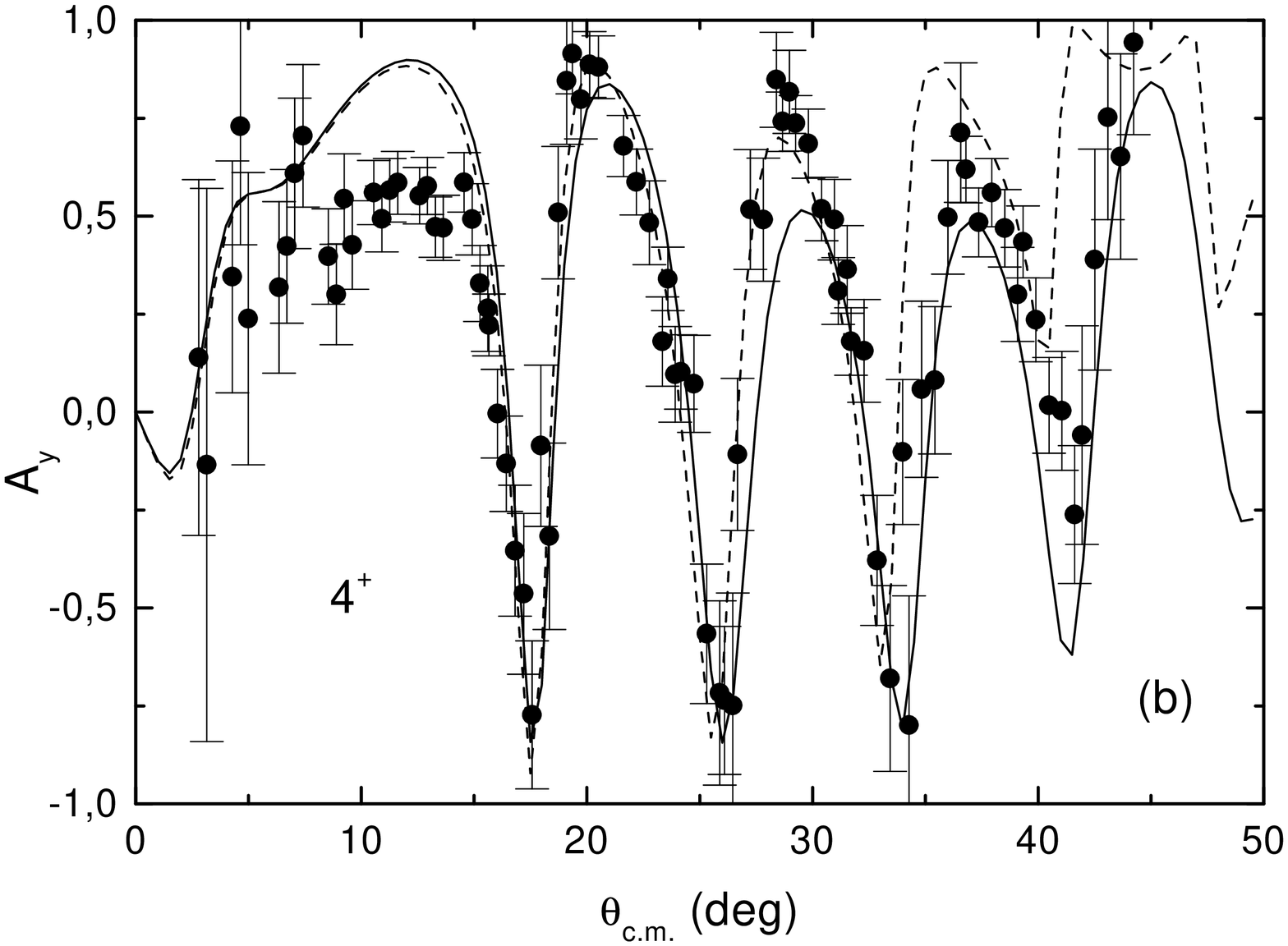}}
\put(0,0) {
\epsfxsize=9.9cm
\epsfysize=6.9cm
\epsfbox{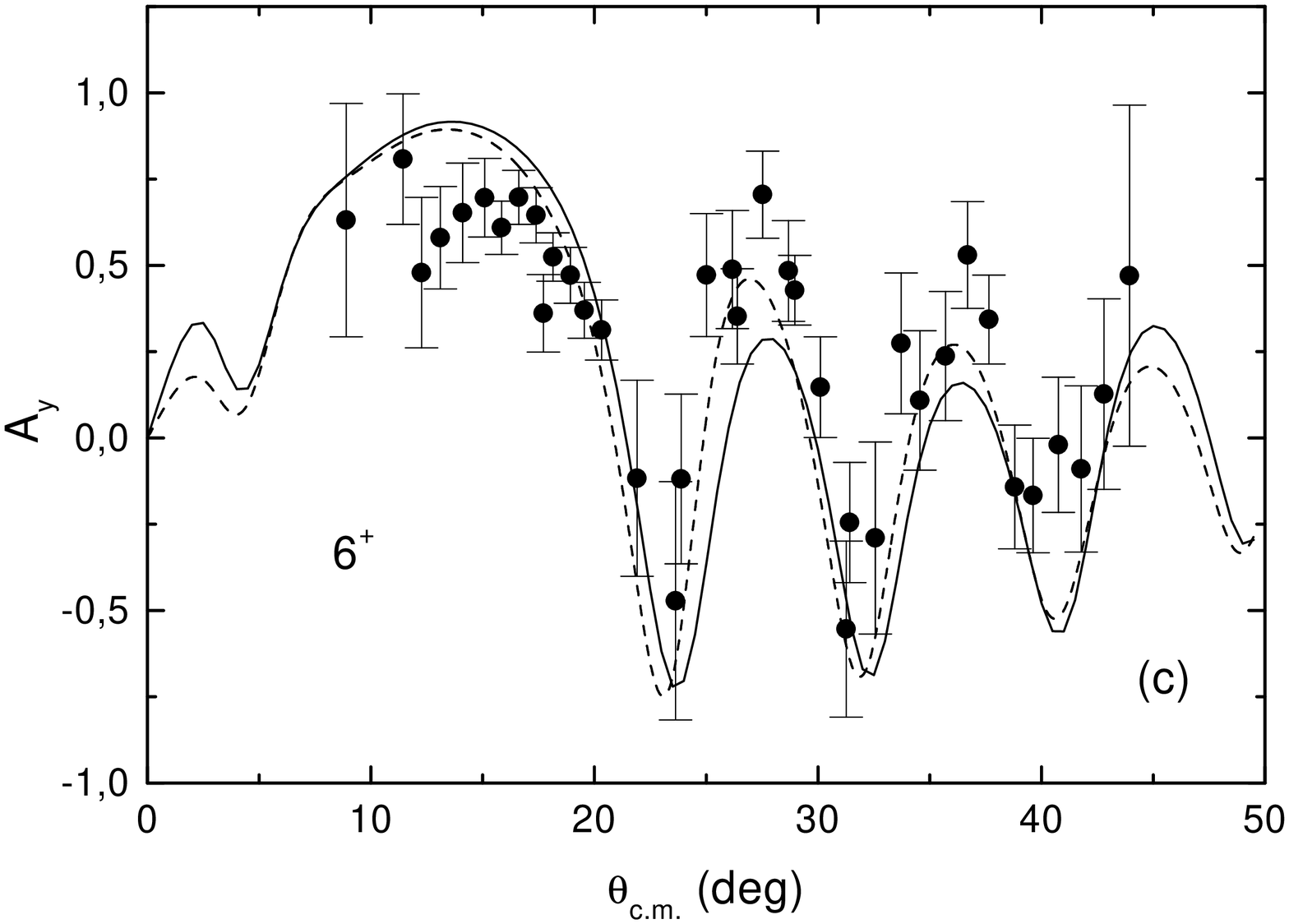}}
\put(250,0) {
\epsfxsize=9.9cm
\epsfysize=6.9cm
\epsfbox{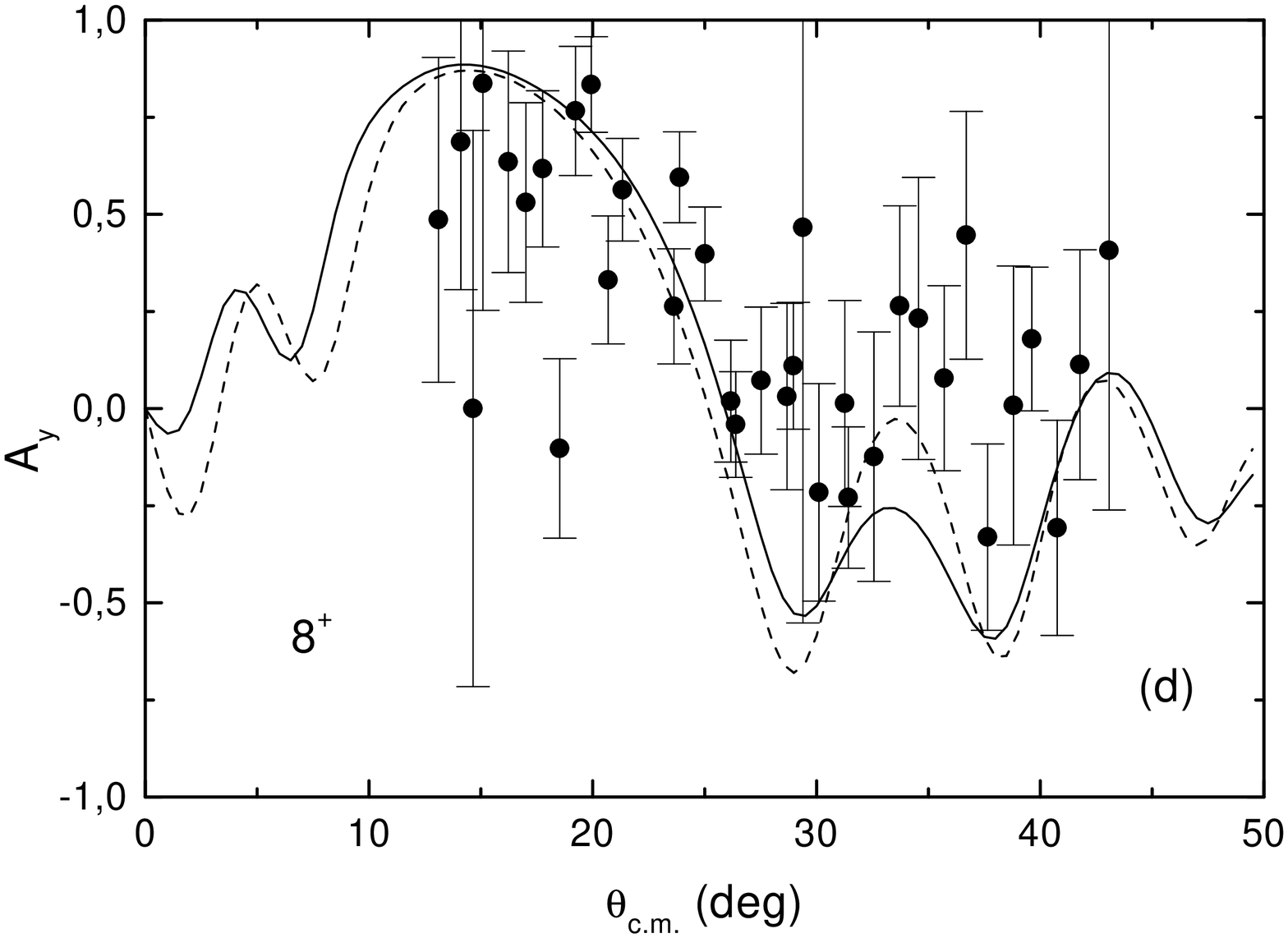}}
\end{picture}

Fig.4. Inelastic analysing powers for $2^+$ (a), $4^+$ (b), $6^+$
(c),
$8^+$ (d) states~\cite{Lee,Lee2} in comparison with DWBA microscopic
folding model calculations  with isoscalar (dash curves) and model
transition densities (full curves).
\end{center}

\end{document}